\documentclass[prl,twocolumn,showpacs,floatfix,aps]{revtex4}
\usepackage{graphicx}

\begin{document}

\preprint{cond-mat/0000000}
\title{Self-energy Effect of Superconducting Energy Gap in Point-Contact
Spectra of MgB$_{2}$}
\author{I. K. Yanson\thanks{%
Corresponding author, e-mail: yanson@ilt.kharkov.ua}, S. I. Beloborod'ko,
Yu. G. Naidyuk}
\affiliation{B.Verkin Institute for Low Temperature Physics and Engineering, National
Academy of Sciences of Ukraine, 47 Lenin Ave., 61103, Kharkiv, Ukraine}
\author{O. V. Dolgov}
\affiliation{Max-Planck Institut f\"{u}r Festk\"{o}rperforschung, Stuttgart, Germany}
\author{A. A. Golubov}
\affiliation{Faculty of Science and Technology, University of Twente, 7500 AE Enschede,
The Netherlands}
\date{\today}

\begin{abstract}
In strong-coupling superconductors with a short electron mean free
path the self-energy effects in the superconducting order
parameter play a major role in the phonon manifestation of the
point-contact spectra at the above-gap
energies. We derive asymptotic expressions of the phonon structure in MgB$%
_{2}$ in the case $eV\gg\Delta $ for tunnel, ballistic, and
diffusive point-contacts and show that these expressions not only
qualitatively, but also semi-quantitatively correspond to the
measurements of the phonon structure in the point-contact spectra
for the $\pi$-band of MgB$_{2}$ $c$-axis oriented thin films.
\end{abstract}

 \maketitle

$Introduction.$

It is commonly accepted that the mechanism of superconductivity in
recently discovered MgB$_{2}$ \cite{Nagamatsu} is due to the
electron-phonon interaction (EPI) \cite{Mazin}. Among the known
$s-p$ metals, the record breaking critical temperature
($T_{c}\approx 40$ K) and the unusual two-band character in its
electronic structure make this compound very interesting for
detailed study. MgB$_{2}$ crystallizes in a hexagonal lattice with
alternating planes of Mg and B. Up to now, there has been no
detailed experimental determination of its EPI spectral function.
The reason is that EPI in MgB$_{2}$ is strongly anisotropic, and
single crystals, well suited for anisotropic measurements by means
of tunneling or point-contact spectroscopy, are still under
development. As for the tunneling spectroscopy, in order to
extract the so-called \textquotedblright driving
force\textquotedblright\ for high $T_{c}$ in this compound caused
by EPI along the $ab$-plane \cite{Choi}, one needs a plane tunnel
junction oriented perpendicular to this direction, which is not an
easy task. On the other hand, the tunneling spectra in the
$c-$direction are supposed to be so weak (see below), that the
precise standard procedure for extracting the EPI function
\cite{Wolf} is difficult to apply. In Ref. \cite{Dolgov} it was
shown that the inversion of the Eliashberg equation for a
multiband superconductor is a mathematically ill-defined problem.
The previous attempt to obtain a solution on polycrystalline
samples contains an uncontrolled
mixture of the two bands with a predominant contribution of the $\pi $-band %
\cite{D'yachenko}. The interaction of the two bands to a great
extent determines the corresponding EPI functions. Their intensity
depends on electron mean free path in each band separately. On the
one hand, in some experiments the two bands can be considered as
being independent of each other, but, on the other hand, both of
them have the same $T_{c}$, despite quite different EPI. The
layered structure and strongly anisotropic EPI allow us to hope
that the peculiarities in MgB$_{2}$ may shed some light on the
mechanism of high-$T_{c}$ superconductivity in copper oxides.

In the previous paper \cite{Yanson} it was shown qualitatively
that
the phonon singularities in the point-contact spectra of the $c$-axis oriented MgB$%
_{2}$ thin films were due to the non-linear dependence of excess current $%
I_{exc}(eV)$ at energies higher than the superconducting energy gap $\Delta $%
. In the present contribution we derive asymptotic expressions for
differential conductance of $N-c-S$ ($c$ stands for
''constriction'') point-contacts and show that quantitatively
(within a factor of order of unity) the calculated point-contact
spectra agree with the experimental data for the $c$-axis
orientation of MgB$_{2}$, where only the $\pi $-band is visible.

$Theory.$

\begin{figure}[tbp]
\includegraphics[width=8cm,angle=0]{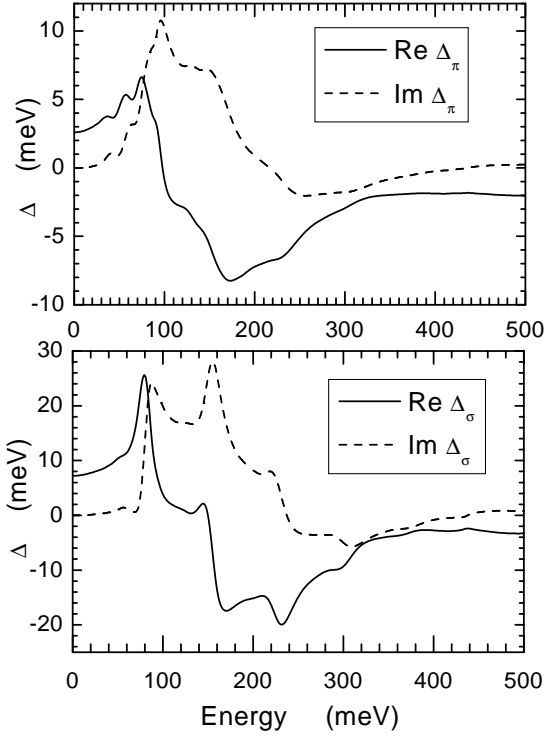}
\caption{The self-energy effects in the complex order parameter of
superconducting MgB$_{2}$. The upper and lower panels are for the $\protect%
\pi $- and the $\protect\sigma $-band, respectively.}
\label{Delta}
\end{figure}

The electronic structure of MgB$_{2}$ consists of two groups of bands \cite%
{Mazin,Choi}: a pair of approximately isotropic 3D $\pi $-bands,
and a pair of strongly anisotropic 2D $\sigma $-bands whose
characteristics are measurable for the point contact oriented
within a few degrees around the $ab $-plane \cite{Brinkman}. As it
was argued in \cite{Mazin1}, the
variation of the superconducting gap inside the $\sigma $- or the $\pi $%
-bands can hardly be observed in real samples due to intraband
impurity scattering. Therefore, superconducting properties of
MgB$_{2}$ can be described by an effective two-band model, where
each group of the $\sigma $- and the $\pi $-bands is characterized
by the corresponding order parameters $\Delta _{\sigma }$ and
$\Delta _{\pi }$. The gap functions $\Delta _{\sigma ,\pi }$ in
MgB$_{2}$ in the Matsubara representation, suitable for the
description of thermodynamic properties, were calculated in Ref.
\cite{Golubov} from the solution of the Eliashberg equations using
the theoretically calculated EPI spectral functions for the
effective two-band model. Here we extend this approach to obtain
the complex gap parameters $\Delta _{\sigma ,\pi }(E)$ in
MgB$_{2}$ as a functions of the real energy $E$. The results are
shown in Fig.~\ref{Delta} and demonstrate the self-energy effects
important for the description of the transport properties across
MgB$_{2}$ point contacts.

If the point contact is clean, then the electron mean free path $l$ is
greater than the size of the constriction $d$, and the current flows in the
ballistic regime. The differential conductance of $S-c-N$ contact at $T=0$
obeys the following expression \cite{Omel'yanchuk}:
\begin{eqnarray}
\left(\frac{dI}{dV}\right)_{bal}(eV)=\frac{1}{R_{N}}\left(
1+\left| \frac{\Delta \left( \epsilon \right) }{\epsilon
+\sqrt{\epsilon ^{2}-\Delta
^{2}\left( \epsilon \right) }}\right| ^{2}\right) \\
{\text{\ }\epsilon =eV},  \nonumber
\end{eqnarray}
where $R_N$ is the normal state resistance.

In the dirty contact $(l \ll d)$, the regime of current flow is
diffusive and the corresponding expression for the differential
conductance follows from formula (21) from Ref.
\cite{Beloborod'ko} for the current-voltage dependence:
\begin{eqnarray}
I(V) =\frac{V}{R_{N}}+\frac{1}{eR_{N}}\int_{0}^{eV}d\epsilon \times \quad
\nonumber \\
\times \left[ \text{Re}\frac{u}{\sqrt{u^{2}-1}}\left( \frac{\text{Re}\text{ }%
\text{Arcsinh} \left( \frac{1}{\sqrt{u^{2}-1}}\right)
}{\text{Re}\left( \frac{1}{\sqrt{u^{2}-1}}\right) }\right)
-1\right] ,
\end{eqnarray}
where $u =\frac{\epsilon }{\Delta \left( \epsilon \right)},$
$\epsilon =eV ,$  which yields
\begin{eqnarray}
\!\!\!\left(\frac{dI}{dV}\right)_{dif}(eV) =\frac{1}{2R_{N}}\ln
\left| \frac{\epsilon+\Delta \left( \epsilon \right) }{\epsilon
-\Delta \left(
\epsilon \right) }\right| \times  \nonumber \\
\times\left[\text{Re}\frac{\epsilon }{\sqrt{\epsilon ^{2}-\Delta^{2}\left(
\epsilon \right) }}/\text{Re}\frac{\Delta (\epsilon )}{\sqrt{\epsilon
^{2}-\Delta ^{2}\left( \epsilon \right) }}\right],\text{ \ } \text{\ \ }%
\epsilon =eV.  \label{diffusive}
\end{eqnarray}

It is well known that in the tunnel regime the differential conductance is
proportional to the quasiparticle density of states \cite{Wolf}:
\begin{equation}
\left(\frac{dI}{dV}\right)_{tun}(eV) =\frac{1}{R_{N}}\text{Re}\frac{%
\epsilon }{\sqrt{\epsilon ^{2}-\Delta ^{2}\left( \epsilon \right) }},\text{ }
\\
\text{\ \ \ \ }\epsilon =eV.  \label{tunnel}
\end{equation}
We assume $T=0$ in these formulae, but they can be generalized to
a finite temperature as well.

These formulae acquire a very simple form in the limit $eV\gg\Delta $. For
tunnel, ballistic and diffuse point contacts they are as follows:
\begin{equation}
\left(\frac{dI}{dV}\right)_{tun}(eV)\approx \frac{1}{R_{N}}\left[ 1+\frac{%
\text{Re}^{2}\Delta \left( \epsilon \right) }{2\epsilon ^{2}}-\frac{\text{Im}%
^{2}\Delta \left( \epsilon \right) }{2\epsilon ^{2}}\right] ,\text{ \ \ \ }
\label{asymptun}
\end{equation}
\begin{equation}
\left(\frac{dI}{dV}\right)_{bal}(eV)\approx \frac{1}{R_{N}}\left[ 1+%
\frac{\text{Re}^{2}\Delta \left( \epsilon \right) }{4\epsilon ^{2}}+\frac{%
\text{Im}^{2}\Delta \left( \epsilon \right) }{4\epsilon ^{2}}\right] ,\text{
\ \ \ \ }  \label{asympbal}
\end{equation}
and
\begin{eqnarray}
\left(\frac{dI}{dV}\right)_{dif}(eV) \approx \frac{1}{R_{N}}\left[ 1+%
\frac{\text{Re}^{2}\Delta \left( \epsilon \right) }{3\epsilon ^{2}}\right] ,%
\text{ \ \ \ \ }  \label{asympdif}
\end{eqnarray}
where $\epsilon =eV$, respectively. Thus, the
$dI/dV$-characteristics of any point contact (tunnel, ballistic,
diffusive) are roughly similar below 80~meV, where the contribution of Im$%
\Delta(\epsilon)$ is negligible.

\begin{figure}[tbp]
\includegraphics[width=8cm,angle=0]{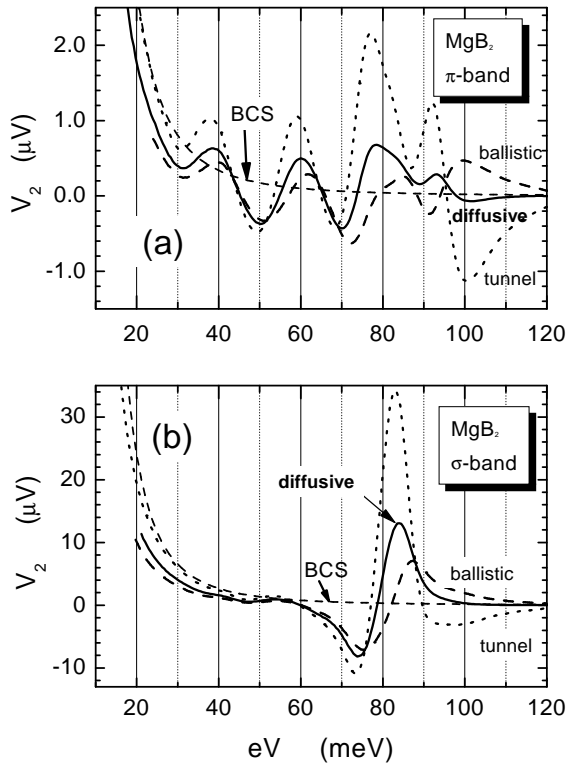}
\caption{The calculated second harmonic signal $V_{2}(V)$
proportional to the second derivative of the $I-V$-characteristic
of the $N-c-S$ point-contact with a ballistic or a diffusive
regime of current flow (see
text). For comparison, the corresponding curve for the tunnel junction $%
N-I-S $ is shown together with the BCS tunneling density of
states. Calibration of $V_{2}$ is reduced to modulation voltage
$V_{1}$=3 mV.} \label{calcul}
\end{figure}

In order to compare the theoretical predictions with the
experimental spectra we express the calculated spectra as the
second harmonic (i.e., $rms$ voltage, $V_{2}(V)$) of the small
first harmonic modulation voltage $V_{1}$ by expanding the $V(I)$
characteristic  in the Taylor series. Namely, we express the
$V_2$-signal as
\begin{equation}
V_{2}(V)\approx -\frac{V_{1}^{2}}{2\sqrt{2}}\frac{d}{dV}\left( R_{N}\frac{dI%
}{dV}\right) ,  \label{experim}
\end{equation}%
assuming that changes in differential resistance is small and,
correspondingly, the modulation voltage is fixed at $V_{1}$. Any other
spectra can be reduced to the particular value of $V_{1}$ using the equation
(\ref{experim}). In what follows we choose $V_{1}=3$~meV, which is close to
the value used in the experiments.

In Fig.~\ref{calcul} (a), (b) the calculated by formulae (\ref{asymptun}), (%
\ref{asympbal}), (\ref{asympdif}), and (\ref{experim}) second harmonic
voltages $V_{2}(V_{1}=3$~meV$)$ are shown for the $\pi -$ and $\sigma -$%
bands of MgB$_{2}$. These extreme regimes of current flow: tunnel,
ballistic, and diffusive, are calculated using $\Delta (eV)$ functions from
Fig.~\ref{Delta}. One can see a small amplitude (of the order of 1 $\mu V$)
of the phonon structure for the $\pi -$band, and a much stronger $E_{2g}$
phonon mode singularity at about 80 meV for the $\sigma $-band. Up to $%
eV\approx 80$ meV the self-energy structure is determined by Re$\Delta
(\epsilon )$. The amplitude of phonon structure is different, reflecting the
corresponding prefactors $1/2\rightarrow 1/3\rightarrow 1/4$ in the row of
current flow regimes: tunnel~$\rightarrow $~diffusive~$\rightarrow $%
~ballistic. On the other hand, large differences in shape are observed in
the energy range 80$\div $120 meV, where Im$\Delta (\epsilon )$ becomes
appreciable. We shall see below that just at the high energy edge the
experimental curves show some variation due to the uncontrollable changes of
scattering in the contact region.

$Comparison$ $with$ $experiment.$

Since the theoretical spectra are odd with regard to the changes
of the $dc$
voltage sign, we extract the odd part of the raw data as shown in Fig.~\ref%
{odd}:
\begin{equation}
V_{2}^{odd}=\frac{1}{2}\left[ V_{2}(eV)-V_{2}(-eV)\right] .
\end{equation}
\begin{figure}[tbp]
\includegraphics[width=8cm,angle=0]{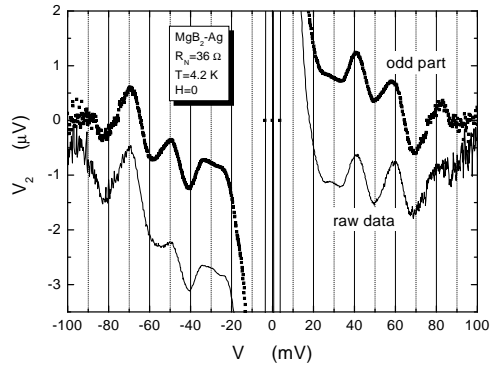}
\caption{The extraction of an odd part of the experimental point-contact
spectra shown for the middle curve in Fig. 4 (a). The thinner curve shows
the raw data shifted vertically down for clarity.}
\label{odd}
\end{figure}

\begin{figure}[tbp]
\includegraphics[width=8cm,angle=0]{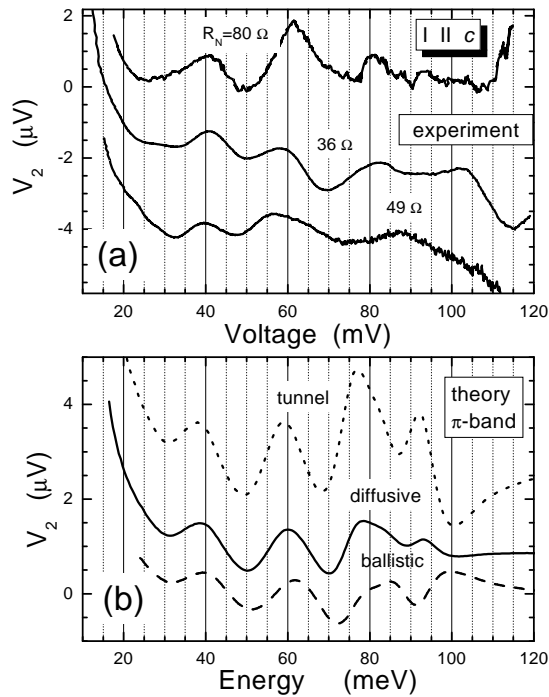}
\caption{Comparison of three different experimental point-contact spectra
(panel $(a)$) in the $c$-direction (Ref. \protect\cite{Yanson}) with the
theoretical ones (panel $(b)$) calculated by asymptotic formulae~( \ref%
{asymptun}), (\ref{asympbal}), (\ref{asympdif}), and
(\ref{experim}) based on superconducting order parameter $\Delta
(\protect\epsilon )$ from Fig.~\ref{Delta}. All the curves are
displaced vertically for clarity and reduced to modulation voltage
$V_{1}=3$~mV. In panel (a), the normal state resistances 49 and 36
$\Omega$ correspond to zero-bias $R_0$= 45 and
43 $\Omega$, respectively, being the same as in Fig. 2 (b) of Ref. %
\protect\cite{Yanson}. For $R_N=80$ $\Omega$ $R_0$=150 $\Omega$. The
experimental curves are smoothed by several mV not disturbing the phonon
structure. $T$=4.2 K, $H$=0.}
\label{compare}
\end{figure}

In Fig.~\ref{compare} we compare 3 odd parts of the experimental spectra for
different point-contacts which are perpendicular to the surface of the $c$%
-oriented thin film \cite{Kang}. The superconducting energy gaps $\Delta
_{0} $ measured by the Andreev reflection spectroscopy from $dV/dI$%
-characteristic lie in the range 2.1$\div $2.6 meV. These $dV/dI$-curves
show almost no traces connected with the large gap at about 7~meV. Hence,
for this orientation we probe only the $\pi $-band \cite{Brinkman}. We
compare the experimental curves (Fig.~\ref{compare} (a)) with the $\pi $%
-band calculated spectra (Fig.~\ref{compare} (b)), since in the experiments
the regime of current flow is among the extremes calculated by the theory.
One can see that not only the shape of the experimental spectra corresponds
well to the theoretical ones (with minor deviations), but also the amplitude
of the structure has a proper order of magnitude ($\sim $ 0.1~\% of $R_{N}$%
). From the $dV/dI(V)$Andreev reflection spectra (see the caption
in Fig. \ref{Compare}) one can suggest that our contacts are
closer to the tunnel regime in the series of the experimental
curves with resistances $49\rightarrow 36\rightarrow 80$~$\Omega
$. To make the estimation more quantitative, one needs an
interpolation formula similar to the finite barrier parameter $Z$
in the BTK-theory \cite{BTK}.

The question arises as to why we did not observe a much stronger self-energy
S-type-structure in Ref. \cite{Naidyuk}, where the point-contact spectrum of
MgB$_2$ single crystal was measured along $ab$-plane? The answer is that in
that case the nonequilibrium phonon generation is so strong that
superconductivity in the contact region is destroyed, and correspondingly,
the excess current dramatically decreases around the $E_{2g}$ phonon mode
energy. This leads to a large maximum in the $dV/dI(V)$ characteristic and
the N-shape singularity in $V_{2}(V)$. The maximum can be easily seen in the
$V_{1}(V)$-characteristic (note, for example, the inset in Fig. 1 of Ref. %
\cite{Naidyuk}). Normally, this destruction of superconductivity masks the
self-energy structure effectively. It can be identified by the strong
dependence of energy position on external parameters: magnetic field and
temperature.

$Summary.$

In conclusion, we have derived the simple asymptotic formulae for
the order parameter self-energy effects in the superconducting
point-contact. They can be used in standard programs
\cite{Rowell,Donetsk} to solve the Eliashberg equations
\cite{Wolf} for quantitatively derivation of
electron-phonon-interaction spectral function providing the
structure of point-contact is established. In MgB$_{2}$, we
applied them in $c$-direction for obtaining point-contact spectra
in $\pi $-band. The close similarity
between the calculated and measured point-contact spectra in the $c$%
-direction manifests the validity of the calculated EPI spectral
function in the $\pi $-band in Ref. \cite{Golubov}.

$Acknowledgements.$

The authors are grateful to N.~L.~Bobrov and V.~V.~Fisun for their
collaboration in the MgB$_{2}$ investigation. O.V.D. and A.A.G.
thank S.~V.~Shulga for the help in numerical calculations. The
work in Ukraine was carried out in part by the State Foundation of
Fundamental Research under Grant $\Phi $7/528-2001.

\end{document}